\documentclass{article}

\PassOptionsToPackage{numbers,compress}{natbib}

\usepackage{amssymb}

\usepackage[final]{neurips_2022_ml4ps}

\usepackage[utf8]{inputenc} 
\usepackage[T1]{fontenc}    
\usepackage{hyperref}       
\usepackage{url}            
\usepackage{booktabs}       
\usepackage{amsfonts}       
\usepackage{nicefrac}       
\usepackage{microtype}      
\usepackage{xcolor}         
\usepackage{listings}
\usepackage{pgfplots}
\usepackage{tikz}
\usepackage{amsmath}
\usepackage{amsthm}
\usepackage{multirow}
\usepackage{mathtools}
\usepackage{enumitem}
\usetikzlibrary{external}

\usepackage{subcaption}

\definecolor{codegreen}{rgb}{0,0.6,0}
\definecolor{codegray}{rgb}{0.5,0.5,0.5}
\definecolor{codepurple}{rgb}{0.58,0,0.82}
\definecolor{backcolour}{rgb}{0.95,0.95,0.92}

\lstdefinestyle{mystyle}{
  backgroundcolor=\color{backcolour},   commentstyle=\color{codegreen},
  keywordstyle=\color{magenta},
  numberstyle=\tiny\color{codegray},
  stringstyle=\color{codepurple},
  basicstyle=\ttfamily\footnotesize,
  breakatwhitespace=false,         
  breaklines=true,                 
  captionpos=b,                    
  keepspaces=true,                 
  numbers=left,                    
  numbersep=5pt,                  
  showspaces=false,                
  showstringspaces=false,
  showtabs=false,                  
  tabsize=2
}

\makeatletter
\newcommand\nocaption{%
    \renewcommand\p@subfigure{}
    \renewcommand\thesubfigure{\thefigure\alph{subfigure}}
}
\makeatother

\title{Deformations of Boltzmann Distributions}

\author{%
  Bálint Máté \\
  University of Geneva\\
  \texttt{balint.mate@unige.ch} \\
  \And
  François Fleuret \\
  University of Geneva\\
  \texttt{francois.fleuret@unige.ch}\\
}

\newcommand{\divg}{{\ensuremath{\operatorname{div}\,}}}

\newcommand{\1}{{\ensuremath{t_1}}}
\newcommand{\2}{{\ensuremath{t_2}}}
\newcommand\norm[1]{\left\lVert#1\right\rVert}

\begin{document}

\maketitle

\begin{abstract}

Consider a one-parameter family of Boltzmann distributions $p_t(x) = \tfrac{1}{Z_t}e^{-S_t(x)}$. This work studies the problem of sampling from $p_{t_0}$ by first sampling from $p_{t_1}$ and then applying a transformation $\Psi_{t_1}^{t_0}$ so that the transformed samples follow $p_{t_0}$. We derive an equation relating $\Psi$ and the  corresponding family of unnormalized log-likelihoods $S_t$. The utility of this idea is demonstrated on the $\phi^4$ lattice field theory by extending its defining action $S_0$ to a family of actions $S_t$ and finding a $\tau$ such that normalizing flows perform better at learning the Boltzmann distribution $p_\tau$ than at learning $p_0$.

\end{abstract}

\section{Introduction}
\begin{figure*}[t!]
    \centering
    \begin{subfigure}[t]{0.48\textwidth}
        \centering
        \includegraphics[width=\textwidth]{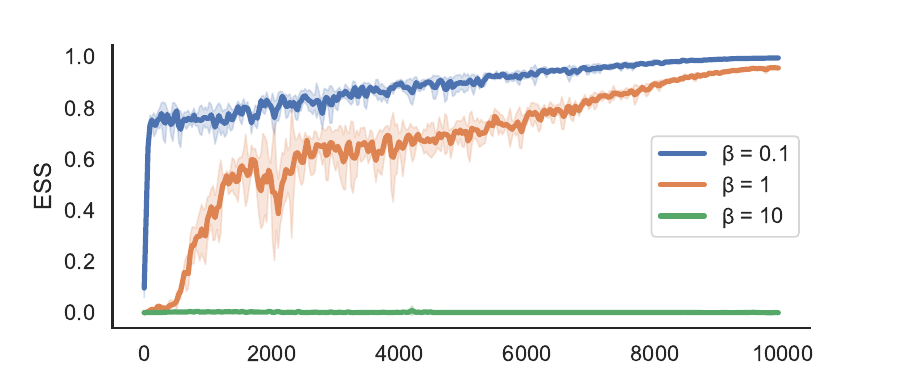}
    \end{subfigure}%
    \begin{subfigure}[t]{0.48\textwidth}
        \centering
        \includegraphics[width=\textwidth]{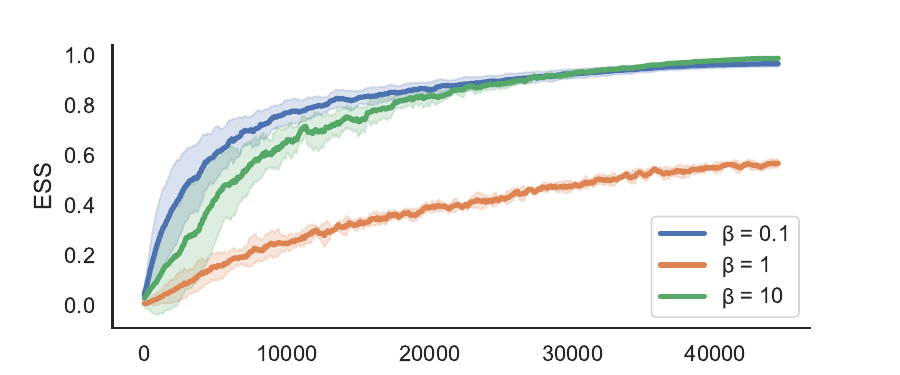}
    \end{subfigure}
    \begin{subfigure}[t]{0.48\textwidth}
        \centering
        \includegraphics[width=\textwidth]{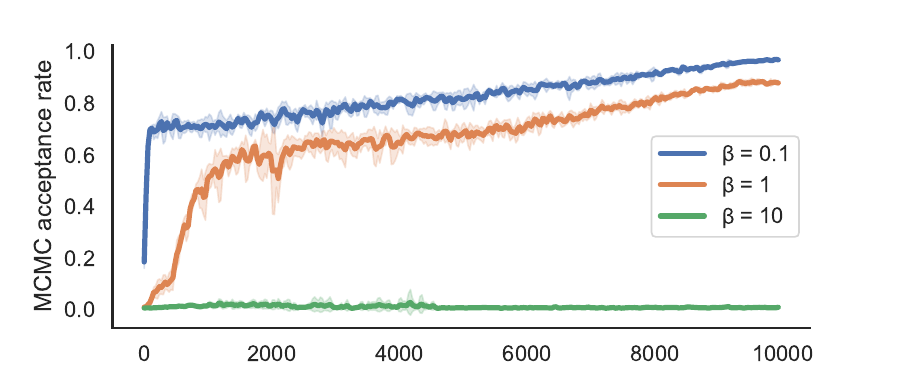}
    \end{subfigure}
    \begin{subfigure}[t]{0.48\textwidth}
        \centering
        \includegraphics[width=\textwidth]{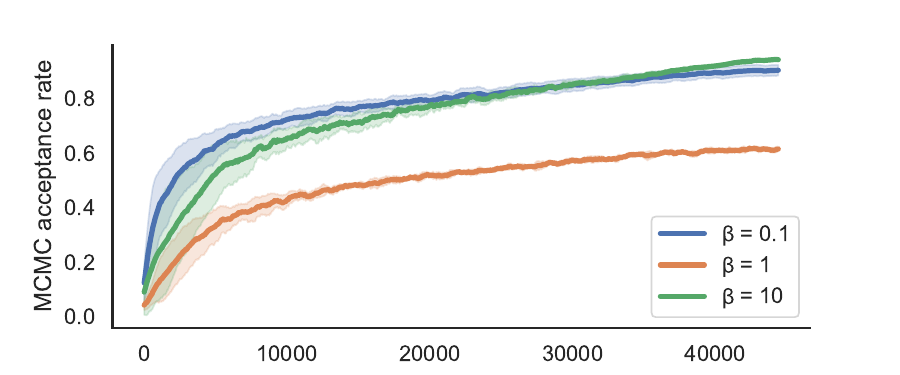}
    \end{subfigure}
    \caption{Sensitivity of the training to the inverse temperature $\beta$ on a $12\times 12$ lattice. We trained the continous normalizing flow of \citet{cnf2} (left) and a RealNVP (right). In both cases we used the action of the $\phi^4$ theory with different values of $\beta \in \{0.1,1,10\}$ and with $m^2$ and $\lambda$  values same as in the work of  \citet{cnf2}. The $x$-axis represents the number of training steps, while the $y$-axis represents the performance metrics. Mean and standard deviation  over 5 runs are shown.}
    \label{fig:beta}
\end{figure*}

Sampling from unnormalized densities has been studied by many due to its relevance for the sciences ~\citep{PhysRevD.103.074504,flow1,Albergo_2021,flow2,flow3,cnf1,cnf2,BoltzmannGenerators,eqFlows,Nicoli_2020,nicoli2021estimation}.
The problem can be summarized as follows. Given an unnormalized log-density $S: \mathbb{R}^n \rightarrow \mathbb{R}$ can we efficiently generate samples from the probability density $p(x) = \frac{1}{Z}e^{-S(x)}$? In particular, there are no samples given, all we have is the ability to evaluate $S$ for any  sample candidate. A popular technique for attacking this problem is to use a normalizing flow  to parametrise a  distribution $q_\theta$ and optimize the parameters $\theta$ to minimize the reverse KL divergence 
\begin{equation}
    KL[q_\theta,p]=\mathbb{E}_{x\sim q_\theta}(\log q_\theta(x)-\log p(x))=\mathbb{E}_{x\sim q_\theta}(\log q_\theta(x)+S(x))+Z.
\end{equation}
As a motivating example for this paper, let $S$ be the defining action of the lattice $\phi^4$ theory (See \S \ref{sec:phi4} for details), and consider the family of distributions $p_\beta(x) \propto e^{-\beta S(x)}$ parametrized by $\beta \in \mathbb{R}^+$.
In terms of statistical physics $\beta$ corresponds to the inverse temperature and controls how ordered the given system is. As seen in Fig.~\ref{fig:beta} the performance of a normalizing flow is sensitive to the parameter $\beta$. Continuous normalizing flows converge faster at higher temperatures. Somewhat surprisingly, the RealNVP architecture converges faster both at $\beta = .1$ and $\beta =10$ than at $\beta =1$. A possible explanation is that both the smoother distribution ($\beta=0.1$) and the more localized ($\beta = 10$) is easier to learn than a combination of these characteristics at $\beta=1$.

Motivated by this observation, this paper studies the following problem. Suppose we are given a family of actions $S_t(x)$, parametrised by $t$, inducing a family of distributions, $p_t(x) \propto e^{- S_t(x)}$. We are interested in how these distributions are related for different values of $t$.
From a practical viewpoint, the goal is to sample from $p_{t_2}$ by first sampling from $p_{t_1}$ and making the samples flow along some vector field $V_t$. We will refer to this $V_t$ as the deformation field or transport field.

The contributions of this paper can be summarized as:
\begin{itemize}
    \item In \S \ref{sec:def_eq} we derive a PDE, the deformation equation,  that translates between infinitesimal deformations of actions and the infinitesimal deformations of the Boltzmann distributions they induce. 
    \item In \S \ref{sec:phi4} we put the theory into practice in the case of a simple deformation of the lattice $\phi^4$ theory and show that it improves the performance of normalizing flows.
\end{itemize}

\begin{figure*}[h!]
    \centering
    \includegraphics[width=\textwidth,trim={0 13cm 0 13cm},clip]{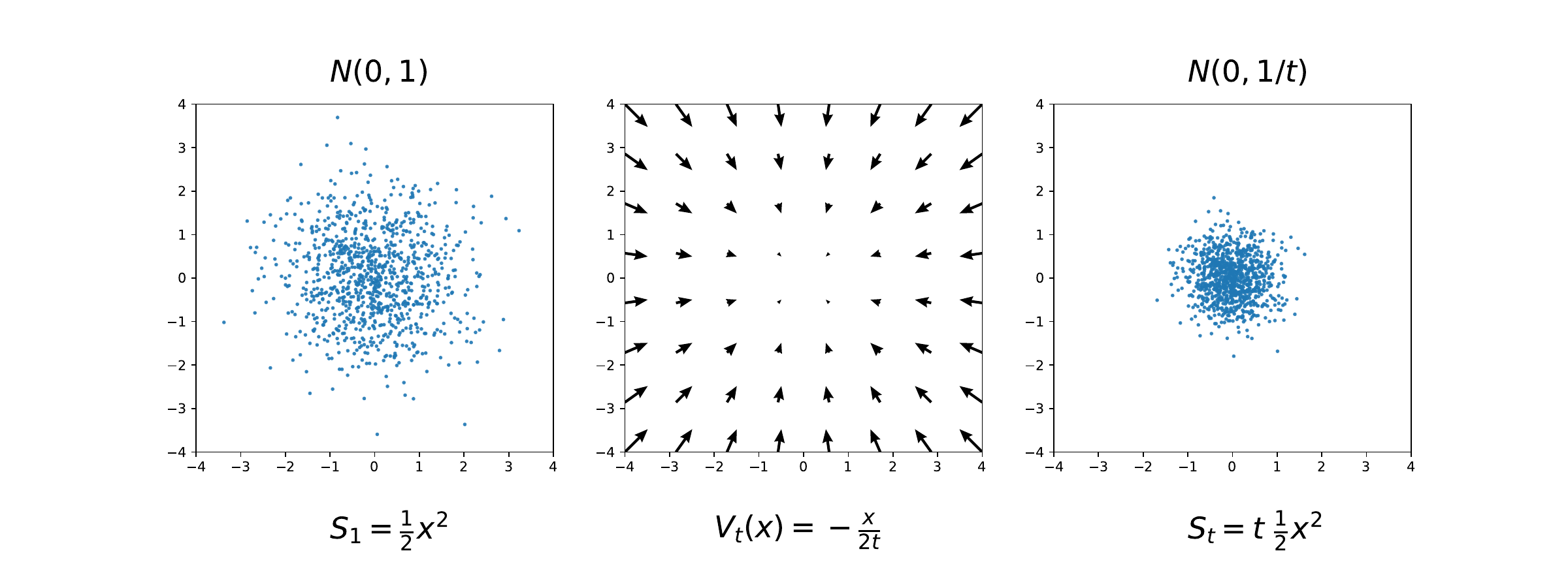}
    \caption{
     Deformation of a two-dimensional standard Gaussian into a family of isotropic Gaussians with different variances. Samples from a two-dimensional standard Gaussian (left). The deformation field (center). The samples in the left plot flowing along the deformation field from $1$ to $t$ follow a Gaussian centered at the origin with covariance $1/t$ (right).}
    \label{fig:gaussian_example_plot}
 \end{figure*}

\section{Background}

\paragraph{Change of variables}

Let  $p_0(z)$ be a probability density on $\mathbb{R}^n$ and $\Psi:\mathbb{R}^n \rightarrow\mathbb{R}^n$ a diffeomorphism. Pushing the the density forward along $\Psi$ induces a new  probability density $p$ implicitly defined by
\begin{equation}
    \label{eq:change_of_variables}
    \log p_0(z)=  \log p(\Psi z) + \log | \det J_{\Psi}(z)|.
\end{equation}
The term $\log | \det J_{\Psi}( z)|$ measures how much the function $\Psi$ expands volume locally at $z$. 
\paragraph{Continous change of variables}
Suppose $V_t$ is a time-dependent vector field. Let $\Psi_\tau$ denote the diffeomorphism of following the trajectories of $V_t$ from $0$ to $\tau$.  This family of diffeomorphisms generates a one-parameter family of densities $p_\tau$. 
The amount of volume expansion a particle experiences along this trajectory of $\Psi_\tau$ is $\int_{0}^{\tau} \divg{V_t}(\Psi_{t}z) dt$. The log-likelihoods are then related by 
\begin{equation}
    \label{eq:cont_change_of_variables}
     \log p_0(z) = \log p_\tau(\Psi_{\tau} z)+ \int_{0}^\tau \divg{V_t}(\Psi_{t}z) dt.
\end{equation}
\newpage
\paragraph{Normalizing flows}
Normalizing flows\citep{tabak2013family,dinh2016density,kingma2018glow,durkan2019neural} (continous normalizing flows \citep{neuralODE}, respectively) parametrize a subset of the space of all distributions on $\mathbb{R}^n$. They do this by first fixing a base density $p_0$ and using a neural network that parametrizes the transformation $\Psi$ (the vector field $V_t$, respectively). The change of variables formula in Eq. \ref{eq:change_of_variables} (Eq.~\ref{eq:cont_change_of_variables}, respectively) is then applied to compute the distribution induced by $\Psi$ ($V_t$, respectively).

\paragraph{Performance metrics} To evaluate the performance of normalizing flows, we employ the following two metrics.
First, let $q_\theta$ be the distribution parametrized by a normalizing flow and $p\propto e^{-S}$ the target distribution. Given a batch of samples $x_i$, the effective sample size can be computed as
\begin{equation}
    ESS = \frac{\left(\tfrac{1}{N} \sum_i p[x_i]/q_\theta[x_i]\right)^2}
    {\tfrac{1}{N} \sum_i (p[x_i]/q_\theta[x_i])^2 }.
\end{equation}
We now think of the samples as being generated sequentially in a MCMC manner. We always accept the first sample $x_0$, and each of the following samples are accepted with probability
\begin{equation}
    p_{accept}(x^i|x^{i-1})= \operatorname{min} \left(1, \frac{q_\theta[x^{i-1}]}{p[x^{i-1}]} \frac{p[x^{i}]}{q_\theta[x^{i}]}\right).
\end{equation}
Note that all ratios $p[x]/q_\theta[x]$ are invariant to change of coordinates because both the numerator and denominator get multiplied with determinant of the same Jacobian. This in turn implies that both the ESS and the MCMC acceptance probability are invariant under a change of coordinates.

\paragraph{Boltzmann distributions}
Let $S:\mathbb{R}^n \rightarrow \mathbb{R}$ be a function with a finite normalizing constant $Z= \int e^{- S(x)} d^n x$. Physically speaking, $S$ might be the action of some physical theory or the energy associated to configurations of some system. Either way, $S$ induces a Boltzmann distribution over the configurations $x\in \mathbb{R}^n$
\begin{equation}
    \label{gibbs}
    p(x) = \frac{1}{Z}e^{- S(x)}.
\end{equation}
Conversely, given a probability density function $p : \mathbb{R}^n \rightarrow \mathbb{R}_{+,0}$ the corresponding action can be recovered up to a constant $S = - \log p - \log Z$ and Eqs. \ref{eq:change_of_variables} and \ref{eq:cont_change_of_variables} can be interpreted as equations relating (the actions and normalizing constants of) two physical theories over the same configuration space.

\section{The deformation equation}
\label{sec:def_eq}
In what follows, we will use the symbols $\partial_tf$ and $\nabla f$ to denote the derivatives with respect to the time parameter and the gradient over the remaining $n$, spatial components of a time-dependent scalar valued function $f: \mathbb{R}\times\mathbb{R}^n \rightarrow \mathbb{R}$.

The Boltzmann distribution of Eq. ~\ref{gibbs} can be equivalently defined by setting the ratios of likelihoods between all $x$ and $y$ to
\begin{equation}
    \label{gibbs2}
    \frac{p(x)}{p(y)} =  e^{S(y)-S(x)}.
\end{equation}
 Our aim is then to construct the time-dependent vector field $V_t$ such that flowing along $V_t$ from $\1$ to $\2$ transforms between the two densities 
\begin{equation}
      p_{\1}(x)=p_{\2}(\Psi_\1^\2x)\mathcal L_{\Psi_\1^\2}(x),
\end{equation}
where $\Psi_\1^\2$ denotes the operation of "flowing along $V_t$ from $\1$ to $\2$"  and $\mathcal L_{\Psi_\1^\2}$ is the likelihood contribution of $\Psi_\1^\2$. In terms of Eq.~\ref{gibbs2}, we are looking for ${\Psi}$ such that for any pair of $x,y$ we have
\begin{equation}
    \frac{p_{\1}(x)}{p_{\1}(y)}=
    \frac{p_{\2}(\Psi x)}{p_{\2}(\Psi y)}\frac{\mathcal L_\Psi(x)}{\mathcal L_\Psi(y)},
\end{equation}
where we dropped the indices from $\Psi$ for readability.
For Boltzmann distributions, this expression is equivalent to
\begin{equation}
    e^{S_\1(y)-S_\1(x)}= e^{S_\2(\Psi y)-S_\2(\Psi x)} \frac{\mathcal L_\Psi(x)}{\mathcal L_\Psi(y)}.
\end{equation}
Taking the logarithm and rearranging
\begin{equation}
 S_\1(x)- S_\2(\Psi x)+ \log \mathcal L_\Psi(x)=  S_\1(y) -  S_\2(\Psi y)  + \log \mathcal L_\Psi(y).
\end{equation}
This is satisfied if and only if the expression (with reinserted indices on $\Psi$)
\begin{equation}
    \label{constant_f}
    f_\1^\2(x)= S_\1(x)- S_\2(\Psi_\1^\2 x)+ \log \mathcal L_{\Psi_\1^\2}(x),
\end{equation} is independent of $x$. Which happens if and only if $\nabla f_\1^\2(x)$ vanishes for all $x,\1,\2$.
We now construct the vector field $V_t$, the infinitesimal generator of $\Psi$.
Expanding  $\nabla f_\1^\2$ in $\2$ at $\1$ up to first order yields
\begin{align}
    \nabla f_\1^\2(x)&=\underbrace{\nabla f_\1^\2(x)\bigg\rvert_{\2=\1}}_{0} + \underbrace{\partial_\2 \nabla}_{\nabla\partial_\2 } f_\1^\2(x)\bigg\rvert_{\2=\1} \delta t  + \mathcal O (\delta t^2) \\
    &= \nabla\left[ - \partial_t S_\1(x) -S_\1 (\partial_\2 \Psi_\1^\2 x) +  \divg V(x)\right]\bigg\rvert_{\2=\1}\delta t  + \mathcal O (\delta t^2) \\
    &= \nabla\left[ - \partial_t S_\1(x) -\langle \nabla S_\1,V_\1(x)\rangle +  \divg V(x)\right]\delta t  + \mathcal O (\delta t^2).
\end{align}
For our purposes $\nabla f_\1^\2$ must vanish everywhere, therefore $V_t$ must be a solution of
\begin{equation}
    \label{pde_on_V2}
     \nabla \left[\partial_t S_t +\langle \nabla S_t, V_t\rangle- \divg{V_t}\right] = 0 \Leftrightarrow \partial_t S_t +\langle \nabla S_t, V_t\rangle-   \divg{V_t} \thickapprox 0,
\end{equation}

where the symbol $\thickapprox$ means that the two sides only agree up to a spatially constant function.
A closely related equation is derived by \citet[Eq. 6]{pfau2020integrable}. Eq. \ref{pde_on_V2} can also be obtained by taking the gradient of their Eq. 6, removing the $t$-dependent normalizing constant $Z_t$.

Given an infinitesimal deformation $\partial_t S$ of the action $S$,  the vector field $V_t$ generating $\partial_t S$ is obtained by solving Eq. \ref{pde_on_V2} for $V_t$. Conversely, given a vector field $V_t$, the infinitesimal deformation it induces on $S$ is found by solving the same equation for $S_t$.
   \subsection*{Sanity check: 1D Gaussians}
   In general, given a family of actions $S_t$, it is difficult to  solve Eq.~\ref{pde_on_V2} for $V_t$. In the following toy-example we explicitly find a particular solution. We consider one-dimensional centered gaussians that are defined by the action 
   \begin{equation}
   S_t(x)=t\frac{1}{2}x^2 \qquad \Rightarrow \qquad p_t(x) = \frac{1}{Z_t} e^{-\frac{t }{2}x^2} =\mathcal N(\mu=0,\sigma^2=1/t).
   \end{equation}
   Changing $t$ to $t+\epsilon$ changes the distribution from $\mathcal N(0,1/t)$ to $\mathcal N(0,1/(t+\epsilon))$. To map samples from the former to the latter, we can multiply with the ratios of the standard deviations, i.e. apply the map $x \mapsto x\sqrt{\frac{t}{t+\epsilon}}$. The transport field $V_t$ is then given by the derivative of this map with respect to $\epsilon$,

   \begin{equation}
      V_t = \frac{d}{d\epsilon}\left[x\sqrt{\frac{t}{t+\epsilon}}\right]\bigg\rvert_{\epsilon=0}=-\frac{x}{2t}.
   \end{equation}
   This choice for $V_t$ also solves the deformation equation of  the previous section. 
   Indeed, plugging $S_t=t\frac{1}{2}x^2$ and $V_t=-\frac{x}{2t}$ into Eq.~\ref{pde_on_V2},
   \begin{align}
      \partial_x \left[\partial_t S_t+  \langle \nabla S_t,V_t\rangle - \divg V_t \right]
      = \partial_x \left[\frac{1}{2}x^2 - tx \frac{x}{2t}+\frac{1}{2t} \right]
      = 0 .
   \end{align}
This example generalizes to higher dimensional Gaussians, a two dimensional example can be seen in Fig. \ref{fig:gaussian_example_plot}.
\newpage
\section{The $\phi^4$ theory on a 2D lattice}
\label{sec:phi4}
\begin{figure*}[h!]
    \centering
    \begin{subfigure}[t]{0.48\textwidth}
        \centering
        \includegraphics[width=\textwidth]{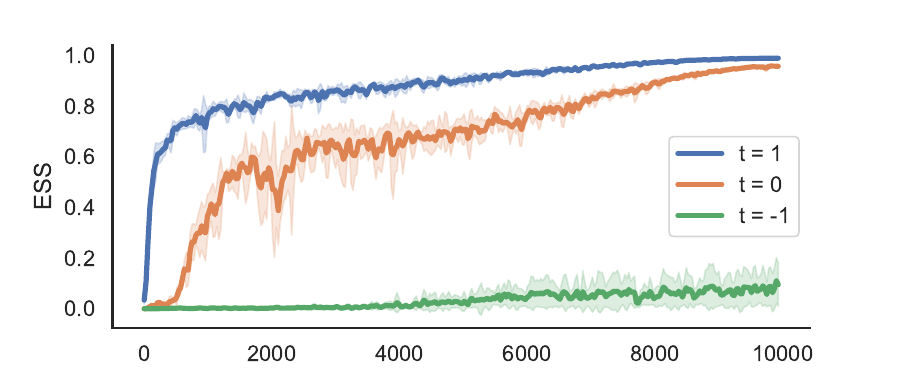}
    \end{subfigure}
    \begin{subfigure}[t]{0.48\textwidth}
        \centering
        \includegraphics[width=\textwidth]{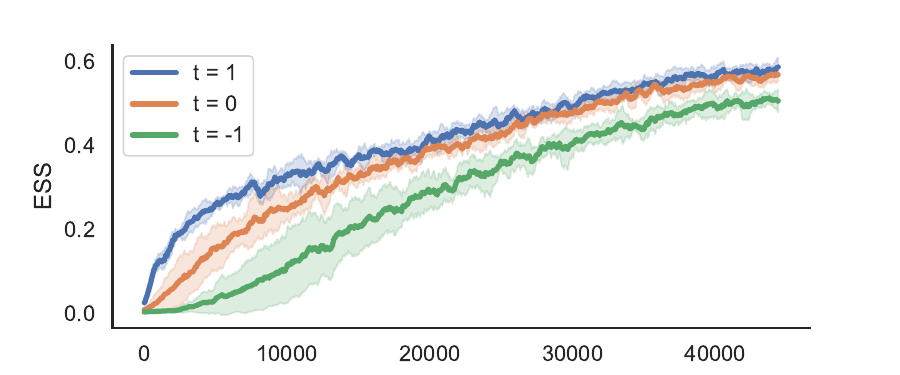}
    \end{subfigure}
    \begin{subfigure}[t]{0.48\textwidth}
        \centering
        \includegraphics[width=\textwidth]{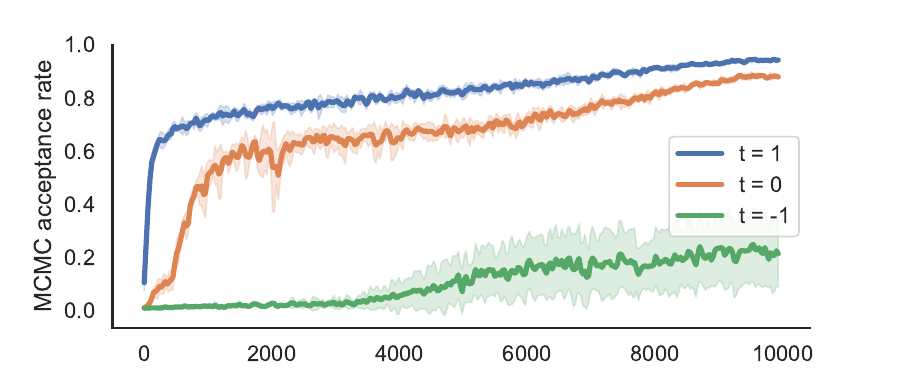}
    \end{subfigure}
    \begin{subfigure}[t]{0.48\textwidth}
        \centering
        \includegraphics[width=\textwidth]{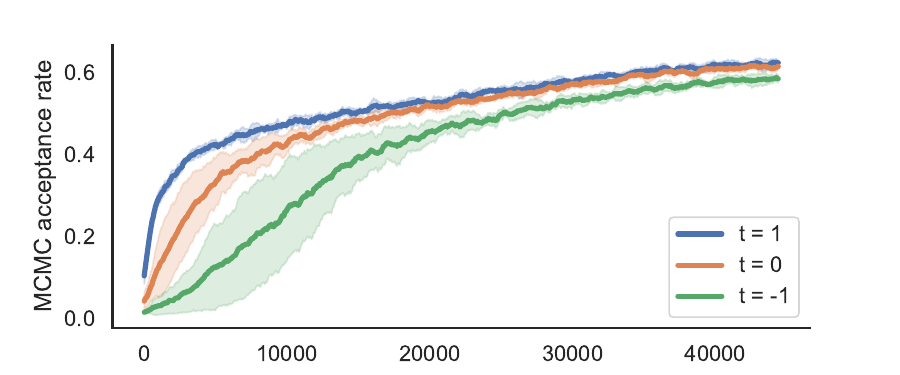}
    \end{subfigure}
    \caption{
     Sensitivity of the training to the deformation parameter $t$ on a $12\times 12$ lattice.  We trained the continous normalizing flow of \citet{cnf2} (left) and a RealNVP  (right). In both cases we used the action given in Eq.~\ref{eq:S_t} with different values of $t \in \{-1,0,1\}$ and with $m^2$ and $\lambda$  values same as in the work of  \citet{cnf2}.  The $x$-axis represents the number of training steps, while the $y$-axis represents the performance metrics. Mean and standard deviation  over 5 runs are shown.}
    \label{fig:defT}
 \end{figure*}

The $\phi^4$ theory is defined by the action $S_0[\phi] = \int \norm{\nabla \phi}^2 + m^2 \phi^2 + \lambda \phi^4\,\, dx$.
After discretizing to a finite lattice $L$, the action becomes
\begin{align}
\sum_{x,y \in L}  \phi(x) \Delta^L_{x,y} \phi(y)+
   \sum_{x\in L}m^2 {\phi}^2(x) +
    \sum_{x\in L}\lambda_0 \phi^4(x)
    =\phi^T\underbrace{\left(\Delta^L + m^2\right)}_{K_0}\phi + \lambda_0 \norm{\phi}_4^4,
\end{align} where $\Delta^L$ is the Laplacian of the lattice $L$. We now deform with one of the simplest non-trivial vector fields, $ V_t(\phi) = V(\phi) = \phi$. As a function, $V$ is just the identity, and the dynamics it induces is a radial expansion centered at the origin.  Plugging it into Eq. ~\ref{pde_on_V2}, 
\begin{align}
\partial_t S_t&= 
\divg V -\langle \nabla S_t, V \rangle  
= \operatorname{dim} \phi -\left \langle 2 K_t \phi+4\lambda_t \phi^3,  \phi  \right \rangle 
\thickapprox -2\phi^T K_t\phi -4\lambda_t  \norm{\phi}_4^4,
\end{align} 
where we assumed that $S_t$ is of the form $S_t= \phi^T K_t\phi + \lambda_t \norm{\phi}_4^4$.
Indeed, setting $K_t = e^{-2t}K_0$ and $\lambda_t = e^{-4t} \lambda_0$ yields a solution,
\begin{align}
    \label{eq:S_t}
 S_t[\phi] =e^{-2t} \big(\phi^T K_0 \phi \big)+ e^{-4 t}\big(\lambda_0 \norm{\phi}_4^4\big).
\end{align} 
An interesting point to note here is the similarity between the family of actions $S_t$ given by Eq.~\ref{eq:S_t} and the family  $S_\beta = \beta S_0$ that motivated this paper. The family $S_\beta$ deforms $S_0$ by scaling the whole action with the same multiplicative constant, while the family $S_t$ deforms $S_0$ by scaling quartic terms quadratically faster than the quadratic terms. 
Fig.~\ref{fig:defT} shows how the deformation in Eq.~\ref{eq:S_t} with a positive value of $t$ improves the performance of normalizing flows.

\section{Conclusion and Impact Statement}
\paragraph*{Conclusion and limitations}We propose a technique that deforms one Boltzmann distribution into an other that is easier to model with normalizing flows. We experimentally support this in the case of the $\phi^4$ lattice field theory. Our experiments indicate that the technique leads to improved ESS and acceptance rates. It is important to note however that good ESS and acceptance rate values does not mean that $q_\theta$ is perfectly matching $p$. For example, if there are several modes of $p$, $q_\theta$  could be missing some of the modes while still obtaining good ESS and acceptance rate values.

\paragraph*{Impact Statement}
Beyond the application in lattice field theory that we presented in this paper, the technique of deforming distributions naturally lends itself to being used for modelling other physical and chemical systems.
The potential impact of the technique introduced in this paper is to allow the above tasks to be solved more efficiently.
In particular, the authors do not see any ethical concerns nor a direct potential of negative impact to society.

\section{Acknowledgement}
The authors acknowledge support from the Swiss National Science Foundation under grant number CRSII5\_193716 -
"Robust Deep Density Models for High-Energy Particle Physics and Solar Flare Analysis (RODEM)".
We further thank Samuel Klein and Jonas Köhler for discussions.
\bibliographystyle{unsrtnat}
\bibliography{bib}
\section*{Checklist}


\begin{enumerate}

\item For all authors...
\begin{enumerate}
  \item Do the main claims made in the abstract and introduction accurately reflect the paper's contributions and scope?
    \answerYes{}
  \item Did you describe the limitations of your work?
    \answerYes{}
  \item Did you discuss any potential negative societal impacts of your work?
    \answerYes{}{}
  \item Have you read the ethics review guidelines and ensured that your paper conforms to them?
    \answerYes{}
\end{enumerate}

\item If you are including theoretical results...
\begin{enumerate}
  \item Did you state the full set of assumptions of all theoretical results?
    \answerYes{}
        \item Did you include complete proofs of all theoretical results?
    \answerYes{}
\end{enumerate}

\item If you ran experiments...
\begin{enumerate}
  \item Did you include the code, data, and instructions needed to reproduce the main experimental results (either in the supplemental material or as a URL)?
    \answerYes{}
  \item Did you specify all the training details (e.g., data splits, hyperparameters, how they were chosen)?
    \answerYes{}
        \item Did you report error bars (e.g., with respect to the random seed after running experiments multiple times)?
    \answerYes{}
        \item Did you include the total amount of compute and the type of resources used (e.g., type of GPUs, internal cluster, or cloud provider)?
    \answerYes{}
\end{enumerate}

\item If you are using existing assets (e.g., code, data, models) or curating/releasing new assets...
\begin{enumerate}
  \item If your work uses existing assets, did you cite the creators?
    \answerNA{}
  \item Did you mention the license of the assets?
    \answerNA{}
  \item Did you include any new assets either in the supplemental material or as a URL?
    \answerNA{}
  \item Did you discuss whether and how consent was obtained from people whose data you're using/curating?
    \answerNA{}
  \item Did you discuss whether the data you are using/curating contains personally identifiable information or offensive content?
    \answerNA{}
\end{enumerate}

\item If you used crowdsourcing or conducted research with human subjects...
\begin{enumerate}
  \item Did you include the full text of instructions given to participants and screenshots, if applicable?
    \answerNA{}
  \item Did you describe any potential participant risks, with links to Institutional Review Board (IRB) approvals, if applicable?
    \answerNA{}
  \item Did you include the estimated hourly wage paid to participants and the total amount spent on participant compensation?
    \answerNA{}
\end{enumerate}

\end{enumerate}

\newpage
\appendix

\section{Experimental details}
\label{app:training_details}
All of our experiments were implemented in PyTorch. We trained all the models with a sample size of 128 at each training step. The learning rate was initialized to $3\times10^{-3}$ and annealed to $0$ following a cosine schedule.

\paragraph*{RealNVP}
The RealNVP architectures  use an alternating checkerboard masking and a 3-layer convolutional network in each layer with kernel size 3 to compute the parameters of the affine transformations. The overall architecture contains 24 such layers.

\paragraph*{Continous NF}
The experiments involving continous normalizing flows used the architecture proposed by \citet{cnf2} with $F=16$ field basis functions and $D=9$ time kernels, and coupling dimensions $F'=8, D'=8$. The 4th order Runge-Kutta solver of the torchdiffeq package \citep{torchdiffeq} with a step size of $0.02$ was used.

\end{document}